\documentclass[graphics,floatfix,nofootinbibprl,margin=2in,nobibnotes,12pt]{article}
\usepackage{jheppub}
\usepackage{graphicx}
\usepackage{cancel}
\usepackage{amssymb}
\usepackage{textcomp}
\usepackage{amsmath}
\usepackage{bm}
\usepackage{units}
\usepackage{feynmp}
\usepackage{times}
\usepackage{hyperref}
\usepackage{epsfig}
\usepackage{color}
\usepackage{float}
\usepackage[T1]{fontenc} 

\newcommand{\TeV}{\textrm{TeV}}
\newcommand{\GeV}{\textrm{GeV}}
\def\lsim{\:\raisebox{-0.5ex}{$\stackrel{\textstyle<}{\sim}$}\:}

\def\lnc{lepton number conservation }

\makeatletter
\renewcommand{\p@subsection}{}
\renewcommand{\p@subsubsection}{}
\makeatother

\title{\boldmath Heavy Higgs Boson Production at Colliders   \\
in the Singlet-Triplet Scotogenic Dark Matter Model
}
\author[a]{Marco Aurelio D\'\i az}
\author[b]{Nicol\'as Rojas}
\author[a]{Sebasti\'an Urrutia-Quiroga}
\author[b]{Jos\'e W.~F.~Valle}
\affiliation[a]{Instituto de F\'\i sica, Pontificia Universidad Cat\'olica de Chile, \\ Avenida Vicu\~na Mackenna 4860, Santiago, Chile.}
\affiliation[b]{Instituto de F\'isica Corpuscular CSIC/Universitat de Valencia, \\ Parc Cient\'ific, calle Catedr\'atico Jos\'e Beltr\'an, 2, E-46980 Paterna, Spain.}

\abstract{
  
  We consider the possibility that the dark matter particle is a
  scalar WIMP messenger associated to neutrino mass generation, made
  stable by the same symmetry responsible for the radiative origin of
  neutrino mass.
  We focus on some of the implications of this proposal as realized
  within the singlet-triplet scotogenic dark matter model.
  We identify parameter sets consistent both with neutrino
  mass and the observed dark matter abundance.
  Finally we characterize the expected phenomenological profile of
  heavy Higgs boson physics at the LHC as well as at future linear
  Colliders.

}

\arxivnumber{1612.06569}
\preprint{IFIC/16-94}

\keywords{Heavy Higgs Production, Collider Production, Dark Matter, Scotogenic Mechanism.}

\begin{document}

\maketitle
\date{}

\section{Introduction}

Particle physics has celebrated two well-deserved Nobel prizes in this
decade.  These were associated to the historic discoveries of the
Higgs boson~\cite{RevModPhys.86.851} at the Large Hadron Collider
(LHC) and of neutrino oscillations using solar and atmospheric
neutrinos~\cite{Kajita:2016cak,McDonald:2016ixn} beautifully confirmed
by laboratory experiments based at reactors and
accelerators~\cite{Forero:2014bxa}.
Though these discoveries culminate decades of thriving searches, by no
means they close our efforts to understand what lies behind these
phenomena~\cite{Valle:2015pba}.
There are, in addition, robust hints from cosmology as to the
incompleteness of the Standard Model. For example, there is a growing
evidence in favor of the existence of non-baryonic dark
matter~\cite{Bertone:2010}, which also indicates the need for new
physics.
Underpinning the elusive nature of dark matter constitutes a most
important challenge in astroparticle physics and modern cosmology.
A popular dark matter candidate is a weakly interacting massive
particle (WIMP), such as the lightest supersymmetric particle,
typically a neutralino, present in supersymmetric models with R parity
conservation. WIMPs can naturally account for the required relic
density thermally and be detectable by nuclear recoil through weak
strength interactions.

Irrespective of the existence of supersymmetry in nature, the problem
of neutrino mass and the explanation of dark matter may have a common
origin, so that the dark matter candidate can be closely associated 
with the mechanism of neutrino mass generation.
For example dark matter stability may be a remnant of the same
symmetry which accounts for the observed pattern of neutrino
oscillations~\cite{Hirsch:2010ru,Boucenna:2012qb,Lavoura:2012cv}.
In this case, the symmetry stabilizing the lightest particle charged
under it, gives rise to a WIMP Dark Matter candidate.
Another possibility, recently proposed in~\cite{Chulia:2016giq}, is
that the dark matter candidate is a scalar WIMP, and its stability
emerges from a symmetry enforcing the Dirac nature of neutrinos.
In another broad class of models the dark matter candidate appears as
a pseudo-Goldstone boson associated to neutrino mass generation, an
old idea~\cite{berezinsky:1993fm} which, in its original realization,
leads to warm decaying dark matter.
In this case dark matter may be detected through subtle effects in the
cosmic microwave background~\cite{Lattanzi:2007ux} as well as
indirectly through X-ray and gamma-ray line searches in the
sky~\cite{Bazzocchi:2008fh,Lattanzi:2013uza}.

In this paper we focus on the case of Scotogenic Models, a beautiful
idea proposed by Ernest Ma~\cite{Ma:2006km} and generalized in
Ref.~\cite{Hirsch:2013ola}. The dark matter can be either scalar or
fermionic and can be interpreted as the radiative messenger neutrino
mass generation. This postulates that the same symmetry which is
responsible for the radiative nature of neutrino masses also
stabilizes dark matter which emerges as the WIMP messenger of neutrino
mass generation. Here we focus on the possibility of scalar scotogenic
DM since the fermionic case has been considered in~\cite{Hirsch:2013ola}.
In Sec.~\ref{sec:gener-about-singl} we describe the main aspects of
the model such as the Yukawa sector and the symmetry responsible for
radiative neutrino masses.
In Sec.~\ref{sec:scalar-sector-model} we describe the scalar sector in
detail, including the identification of the relevant parameter space
consistent, e.g. with neutrino mass and an adequate dark matter relic
density. The resulting scalar mass spectrum, including the dark matter
sector is derived.
The results of our numerical study of Higgs boson production cross
section at the LHC as well as at the proposed CLIC and ILC options are
given in Sec.~\ref{sec:numerical-analysis} . Moreover we give the
predicted charged as well as heavy CP-even Higgs boson decay rates.
Finally, we conclude in Sec.~\ref{sec:conclusions}, and in the
appendix we collect the Feynman rules for the various decays
heavy scalars present in the model.

\section{Generalities About Singlet Triplet Scotogenic Model}
\label{sec:gener-about-singl}
\subsection{The Model and the New Fermions}\label{sec:model}

The model we will work was proposed at \cite{Hirsch:2013ola}. Its
particle content, with the respective quantum numbers, is given at the
table \ref{tab:MatterModel}. As in other Scotogenic Models, the
$\mathbb{Z}_2$ symmetry plays the role of a stabilizing symmetry for
the dark matter candidate and for ensuring radiative generation of
neutrino masses. 
\begin{table}[!h]
\setlength\tabcolsep{0.25cm}
\centering
\begin{tabular}{| c | c | c | c | c | c | c | c |}
\hline
        & \multicolumn{3}{|c|}{Standard Model} &  \multicolumn{2}{|c|}{new fermions}  & \multicolumn{2}{|c|}{new scalars}  \\
        \cline{2-8}
                      &  $L$  &  $e$   & $\phi$  & $\Sigma$ &  $N$   & $\eta$   & $\Omega$ \\
\hline                                                                          
$\mathrm{SU(2)_L}$    &   2   &  1     &    2    &     3    &   1    &    2     &    3     \\
$\mathrm{U(1)_Y}$     &  -1/2 &  -1    &    1/2  &     0    &   0    &    1/2   &    0     \\
$\mathbb{Z}_2$        &   $+$ &  $+$   &   $+$   &    $-$   &  $-$   &  $-$     &  $+$     \\
$L$                   &   1   &   1    &    0    &     0    &   0    &  -1     &    0     \\
\hline
\end{tabular}
\caption{\label{tab:MatterModel}Fields of the STSM and their quantum numbers.}
\end{table}
The most general $\mathrm{SU(3)_c \times SU(2)_L \times U(1)_Y}$,
Lorentz and $\mathbb{Z}_2$ invariant Yukawa Lagrangian is given by
\begin{equation}
- \mathcal{L}_Y = Y_e^{\alpha \beta}\,\overline{L_{\alpha}} \, \phi \, e_{\beta} + Y_{N}^\alpha \, \overline{L_{\alpha}} \, i\sigma_2 \eta^* \, N + Y_{\Sigma}^\alpha \, \overline{L_{\alpha}} \, C\Sigma^\dagger\, i\sigma_2\, \eta^* \, + Y_{\Omega} \, \text{Tr} \left( \overline{\Sigma} \, \Omega \right) \, N + h.c. \label{eq:yukawa}
\end{equation}
where the greek indices stand for lepton generations. The symbol
$\sigma_2$ is the second Pauli matrix in order to conjugate the
$\mathrm{SU(2)_L \times U(1)_Y}$ charges and $C$ is the Lorentz charge
conjugation matrix.  The new Yukawas couplings $Y_{N}^\alpha$ and
$Y_{\Sigma}^\alpha$ parametrize the interactions between new scalars,
new fermions and SM leptons. Moreover, they take part in the neutrino
mass generation mechanism.

On the other hand, the new fermions $N$ and $\Sigma$ have Majorana
mass terms, which are given by:
\begin{equation}
- \mathcal{L}_M = \frac{1}{2} \, M_\Sigma \text{Tr}\left(\, \overline{\Sigma^{c}} \Sigma \right)
+ \frac{1}{2} \, M_N \, \overline{N^{c}} N 
+ h.c. \label{eq:mass}
\end{equation}
where the superindex $c$ stands for the Lorentz charge conjugation in
order to get an invariant Lagrangian. Notice that, after electroweak
symmetry breaking, the Yukawa coupling $Y_\Omega$ mixes the neutral
component of $\Sigma$ with $N$. Thus the mass sector for the neutral
fermions is given by
\begin{equation}
\mathcal{M}_\chi = \left[\begin{array}{cc} M_\Sigma & Y_\Omega v_\Omega \\ 
Y_\Omega v_\Omega & M_N \end{array}\right] \, .
\end{equation}
Hence one can define the mass eigenstates $\zeta_{1,2}$ for the pair
$\left( \Sigma_0,\, N \right)$ as
\begin{equation}
\left[\begin{array}{c}\zeta_1\\ \zeta_2\end{array}\right] = \left[ \begin{array}{cc}
\cos  \theta & \sin \theta \\
-\sin \theta & \cos \theta
\end{array} \right] \, \left[\begin{array}{c} \Sigma^0\\ N\end{array}\right] = { V(\theta)}\left[\begin{array}{c} \Sigma^0\\ N\end{array}\right] \label{eq:fermions},
\end{equation}
With this definition at hand, one can write down also the Yukawas for
these eigenstates as follows:
\begin{equation}
h={\left[\begin{array}{cc} 
\frac{Y_\Sigma^1}{\sqrt{2}}\quad & Y_N^1 \\
\frac{Y_\Sigma^2}{\sqrt{2}}\quad & Y_N^2 \\
\frac{Y_\Sigma^3}{\sqrt{2}}\quad & Y_N^3
\end{array}\right] \cdot V^T(\theta) }\, ,
\label{eq:h-matrix}
\end{equation}
A definition that will prove to be useful for the radiative generation
of neutrino masses.
Notice that, as shown previously~\cite{Hirsch:2013ola}, the lightest
of these states is a suitable WIMP-like dark matter
candidate. Nevertheless, in what follows we will assume a scalar dark
matter candidate made up of the lightest real neutral component of the
$\eta$ field (namely $\eta_R$). This has similar features to Higgs
portal WIMP dark matter, although this model allows tree level
interactions of the field $\eta$ with the SM through leptons as well,
as seen from eq. (\ref{eq:yukawa}). This is a possible way of
distinguishing this model from the Inert Higgs Dark Matter (IHDM)
scenario in which the {\it inert} doublet has no Yukawa interaction
with either leptons or quarks.

\subsection{Radiative Neutrino Masses}
Here we will briefly introduce the mechanism that gives rise to
radiative neutrino mass generation.  This depends also on parameters
in the scalar potential, something that we will discuss in detail
below.
As already mentioned, the $\mathbb{Z}_2$ symmetry is in charge of
ensuring the radiative nature of neutrino masses. The mechanism
forbids the $\eta$ field from getting a vacuum expectation value (vev)
because the $\mathbb{Z}_2$ symmetry is exactly conserved, in such a
way that the Yukawa terms $L \eta \, N$ and $L\, \Sigma\, \eta$ do not
produce neutrino mass at tree level.  Hence neutrino masses are indeed
radiatively induced.

\begin{figure}[!ht]
\begin{center}
\parbox{50mm}{
\begin{fmffile}{radneutmass}
\begin{fmfgraph*}(150,150) 
\fmfleft{i2,i1} \fmfright{o2,o1}
\fmf{dashes}{i1,v1}
\fmfdot{i1}
\fmflabel{$\left< \phi_o \right>$}{i1}
\fmf{dashes}{v1,o1}
\fmfdot{o1}
\fmflabel{$\left< \phi_o \right>$}{o1}
\fmf{dashes,label=$\eta_{R}$ ($\eta_{I}$),left=0.4,tension=0.4}{v2,v1}
\fmf{dashes,label=$\eta_{R}$ ($\eta_{I}$),left=0.4,tension=0.4}{v1,v4}
\fmf{plain_arrow,label=$\nu$,l.s=left}{i2,v2}
\fmf{plain_arrow,tension=0.4}{v2,v3}
\fmfdot{v3}
\fmflabel{$\zeta_\beta$}{v3}
\fmf{plain_arrow,tension=0.4}{v4,v3}
\fmf{plain_arrow,label=$\nu^c$,l.s=right}{o2,v4}
\end{fmfgraph*}
\end{fmffile}}
\bigskip
\caption{Radiative contributions for neutrino masses in the STSM. Notice
that there are two contributions, one from $\eta_R$ and other one from
$\eta_I$.}
\label{fig:radneutmass}
\end{center}
\end{figure}
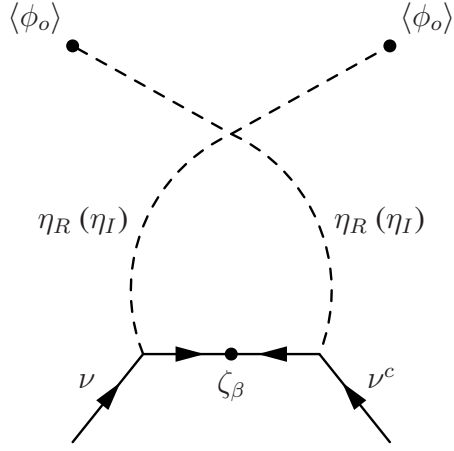

If we include just one of the new fermion fields, either $\Sigma$ or
$N$, one of the columns at the Yukawa matrix (\ref{eq:h-matrix}) gets
removed, so that we will be left with a neutrino mass matrix that has
two zero eigenvalues. However, a setup with one $\Sigma$ and one of
$N$ will lead us to two massive neutrinos and one massless
neutrino. This is enough to account for current oscillation
data~\cite{Forero:2014bxa}. The contributions for radiative neutrino
masses are shown at figure \ref{fig:radneutmass}.

In the figure, we notice the key role of the quartic coupling
$\lambda_{\phi\eta}''$, since it is related to the conservation of
lepton number. While this number has no unique assignment
  within the model, regardless of the choice, it is not a
  conserved quantity in the Lagrangian.  The lepton number assignment
  we will use throughout the paper is shown in the last row at table
  \ref{tab:MatterModel}, in which the $\eta$ field has lepton number
  $-1$. Here, the coupling
$\lambda_{\phi\eta}''\left( \phi^\dagger \eta\right)^2$ breaks lepton
number explicitly, and induces Majorana neutrino masses through the
loop in figure \ref{fig:radneutmass}\footnote{An alternative
  assignment would be when $\Sigma$ and $N$ carry lepton number so
  that the Yukawa interactions at Eq.~(\ref{eq:yukawa}) do conserve
  this number, however, the Majorana masses at Eq.~(\ref{eq:mass}) do
  not.}. Since lepton number gets broken, the resulting effective
neutrino mass matrix is given by:
\begin{equation}
\label{eq:mnuscoto}
{\cal M}_\nu^{\alpha\beta} = \sum_{\sigma=1}^2 \frac{h^{\alpha\sigma}h^{\beta\sigma}}{32\pi^2}
m_{\zeta_\sigma} \left[
\frac{m_{\eta^R}^2 \ln\left({m_{\zeta_\sigma}^2/m_{\eta^R}^2}\right)}{m_{\zeta_\sigma}^2-m_{\eta^R}^2} -
\frac{m_{\eta^I}^2 \ln\left({m_{\zeta_\sigma}^2/m_{\eta^I}^2}\right)}{m_{\zeta_\sigma}^2-m_{\eta^I}^2}
\right]~,
\end{equation}
where $h_{\alpha\beta}$ ($\alpha,\beta = 1,2,3$) are the Yukawa
couplings in the matrix $h$ introduced in Eq.~(\ref{eq:h-matrix}),
$m_{\zeta_\sigma}$ ($\sigma=1,2$) are the masses of the neutral
fermions, and $m_{\eta^{R,I}}$ are the masses of the fields $\eta_R$
and $\eta_I$ respectively. These become degenerate when the coupling
$\lambda_{\phi\eta}'' \to 0$ signaling the \lnc limit where neutrinos
become massless.

As seen above, the structure of the coupling matrix $h$ in
Eq.~(\ref{eq:h-matrix}) is also important for the phenomenology. If we
had just one type of new fermions (either $\Sigma$ or $N$), let's say
$\Sigma$, the structure of the matrix ${\cal M}_\nu^{\alpha\beta}$
will be clealry projective
\begin{equation}
{\cal M}_\nu^{\alpha\beta} \propto 	
\left[ \begin{array}{c}
       Y_{\Sigma}^{1} \\ 
       Y_{\Sigma}^{2} \\
       Y_{\Sigma}^{3}
       \end{array} \right]
\left[ \begin{array}{ccc}
       Y_{\Sigma}^{1} & Y_{\Sigma}^{2} & Y_{\Sigma}^{3}
       \end{array} \right],
\end{equation}
giving rise to only one nonzero eigenvalue, so that one can not
explain neutrino oscillation data~\cite{Forero:2014bxa}.

\section{The Scalar Sector of the Model}
\label{sec:scalar-sector-model}

\subsection{The Scalar Potential}
We now turn our attention to the scalar sector of the model. As seen
in table \ref{tab:MatterModel}, the model includes a SM-like doublet
$\phi$ with hypercharge $Y_\phi=1/2$, a doublet $\eta$ with hypercharge
$Y_\eta=1/2$ and odd $\mathbb{Z}_2$ charge, and a triplet $\Omega$ with
hypercharge $Y_\Omega=0$ and even $\mathbb{Z}_2$ charge, so that its
neutral component can get a nonzero
vev~\cite{Hirsch:2013ola,Merle:2016scw}. Then, the scalar fields of this 
model are written as,
\begin{equation} 
\phi = \left[ \begin{array}{c} \phi^+ \\ \frac{v_\phi+\chi_\phi+i\varphi_\phi}{\sqrt{2}} \end{array} \right]
\,,\qquad
\eta = \left[ \begin{array}{c} \eta^+ \\ \frac{\eta_R+i\eta_I}{\sqrt{2}} \end{array} \right]
\,,\qquad
\Omega = \left[ \begin{array}{cc} \frac{v_\Omega+\chi_\Omega}{\sqrt{2}} & \Omega^+  \\ 
\Omega^- & -\frac{v_\Omega+\chi_\Omega}{\sqrt{2}}  \end{array} \right] \label{eq:scalars}
\end{equation}
{}Then, the corresponding scalar potential is given by,
\begin{eqnarray}
V &=& -m_\phi^2 \phi^\dagger\phi + m_\eta^2 \eta^\dagger\eta 
- \frac{1}{2} m_\Omega^2 {\mathrm{Tr}}(\Omega^\dagger\Omega)
+ \frac{1}{2} \lambda_\phi (\phi^\dagger\phi)^2
+ \frac{1}{2} \lambda_\eta (\eta^\dagger\eta)^2
+ \frac{1}{4} \lambda_\Omega [{\mathrm{Tr}}(\Omega^\dagger\Omega)]^2
\nonumber\\ &&
+ \lambda_{\phi\eta} (\phi^\dagger\phi) (\eta^\dagger\eta)
+ \lambda'_{\phi\eta} (\phi^\dagger\eta) (\eta^\dagger\phi)
+ \frac{1}{2} \lambda''_{\phi\eta} \Big[(\phi^\dagger\eta)^2+(\eta^\dagger\phi)^2\Big]
\nonumber\\ &&
+ \frac{1}{2} \lambda_{\phi\Omega} (\phi^\dagger\phi) {\mathrm{Tr}}(\Omega^\dagger\Omega)
+ \frac{1}{2} \lambda_{\eta\Omega} (\eta^\dagger\eta) {\mathrm{Tr}}(\Omega^\dagger\Omega)
+ \mu_{\phi\Omega} \, \phi^\dagger\Omega\phi + \mu_{\eta\Omega} \, \eta^\dagger\Omega\eta
\end{eqnarray}
Where the mass square parameters $m_\phi^2$, $m_\eta^2$ and
$m_\Omega^2$ are all positive so that $\eta$ is, in principle, unable to get a vev,
so as to prevent breaking the $\mathbb{Z}_2$ symmetry. The quartic
couplings are chosen so that the low energy scalar potential is
bounded from below. Those conditions are given by~\cite{Merle:2016scw}
\begin{eqnarray}
\lambda_\phi \ge 0,\qquad  \lambda_\eta \ge 0,\qquad  \lambda_\Omega &\ge& 0, \\
\lambda_{\phi\eta} + \sqrt{\lambda_\phi \lambda_\eta} \ge 0,\qquad \lambda_{\phi\eta} + \lambda_{\phi\eta}' - |\lambda_{\phi\eta}''| + \sqrt{\lambda_\phi \lambda_\eta} &\ge& 0, \label{eq:la3bounded}\\
\lambda_{\phi\Omega} + \sqrt{2\lambda_\phi \lambda_{\Omega}} \ge 0,\qquad \lambda_{\eta\Omega} + \sqrt{2\lambda_{\eta\Omega} \lambda_\Omega} &\ge& 0 \\
\sqrt{2\lambda_\phi\lambda_\eta\lambda_\Omega} + \lambda_{\phi\eta}\sqrt{2\lambda_\Omega} + \lambda_{\phi\Omega} \sqrt{\lambda_{\eta}} & & \nonumber \\ 
+\, \lambda_{\eta\Omega} \sqrt{\lambda_\phi} + \sqrt{\left(\lambda_{\phi\eta}+\sqrt{\lambda_\phi\lambda_\eta}\right)\left( \lambda_{\phi\Omega} + \sqrt{2\lambda_\phi \lambda_\Omega}\right)\left(\lambda_{\eta\Omega}+\sqrt{2\lambda_\eta\lambda_\Omega}\right)} &\ge& 0.
\end{eqnarray}
These were obtained after using copositivity
conditions~\cite{Kannike:2012pe} inside the scalar potential.  As
shown in \cite{Merle:2016scw}, even if one starts from an adequately
consistent potential at the electroweak scale, one could reach a
situation in which the Lagrangian is no longer invariant under the
$\mathbb{Z}_2$ symmetry at high energies, signalling that this
symmetry could be broken, i.e.  $m_\eta^2\, <\, 0$.

This problem can be avoided, in general, by choosing a not too large
value for the mass parameter $\mu_{\eta\Omega}$, namely, in the order
of $\lsim\, \TeV$.  We stress that this is just a reasonable
prescription and not a full solution. While a more involved scan in
this direction would be required, it lies out of the scope of the
present work (see \cite{Merle:2016scw} for further details).

\subsection{The Scalar Mass Spectrum}
First we notice that after electroweak symmetry breaking, the $\eta$
field has no mixings with other fields, thus the physical masses in
this  $\mathbb{Z}_2$-odd sector are given by
\begin{eqnarray}
m_{\eta^R}^2 &=& m_\eta^2 + 
\frac{1}{2} (\lambda_{\phi\eta}+\lambda'_{\phi\eta}+\lambda''_{\phi\eta}) v_\phi^2
+ \frac{1}{2} \lambda_{\eta\Omega} \, v_\Omega^2
- \frac{1}{\sqrt{2}} \mu_{\eta\Omega} \, v_\Omega \label{eq:etaR} \\
m_{\eta^I}^2 &=& m_\eta^2 + 
\frac{1}{2} (\lambda_{\phi\eta}+\lambda'_{\phi\eta}-\lambda''_{\phi\eta}) v_\phi^2
+ \frac{1}{2} \lambda_{\eta\Omega} \, v_\Omega^2
- \frac{1}{\sqrt{2}} \mu_{\eta\Omega} \, v_\Omega \label{eq:etaI} \\
m_{\eta^+}^2 &=& m_\eta^2 + \frac{1}{2} \lambda_{\phi\eta} \, v_\phi^2
+ \frac{1}{2} \lambda_{\eta\Omega} \, v_\Omega^2
+ \frac{1}{\sqrt{2}} \mu_{\eta\Omega} \, v_\Omega
\end{eqnarray}

Turning to the pseudoscalar $\mathbb{Z}_2$-even sector, it is given
just by $\varphi_\phi$ (the imaginary part of the neutral component of
the Higgs doublet $\phi$ in eq.~(\ref{eq:scalars})), and does not
contain terms from $\eta$ (because it has no vev) or $\Omega$ (because
it has no hypercharge). This means that $\Omega^0$ is real,
so that the $Z$ boson receives a longitudinal component just
from the Higgs-like doublet and not from the triplet.

Something different happens in the charged $\mathbb{Z}_2$-even sector,
since the charged component of the triplet $\Omega$ does mix with the
charged component of the Higgs doublet $\phi^+$. Hence the $W$ boson
longitudinal mode is a linear combination of $\phi^+$ and
$\Omega^{+}$. After solving the tadpole equations for the squared mass
parameters $m_\phi^2$ and $m_\Omega^2$, we obtain the charged scalar
squared mass matrix in the basis $\left(\phi^+ , \Omega^+ \right)^T$
as:
\begin{equation}
M_+^2 = \frac{1}{2\sqrt{2}} \frac{\mu_{\phi\Omega}}{v_\Omega} \left[ \begin{array}{cc} 
4 v_\Omega^2 & 2 v_\phi v_\Omega  \\ 2 v_\phi v_\Omega  & v_\phi^2
\end{array} \right]
\end{equation}
The diagonalization of this mass matrix is performed as
$O_cM_+^2O_c^T={\text{diag}}(m_{G^\pm}^2,m_{H^\pm}^2)$, where the
orthogonal matrix $O_c$ is
\begin{eqnarray}
O_c &=& \left[  
          \begin{array}{cr} 
          c_\beta & - s_\beta \\ 
          s_\beta &   c_\beta
          \end{array}
        \right]
\end{eqnarray}
where the mixing angle is given as
\begin{eqnarray}
c_\beta &=& \frac{v_\phi}{\sqrt{v_\phi^2+4v_\Omega^2}} \,,\qquad
s_\beta \,=\, \frac{2v_\Omega}{\sqrt{v_\phi^2+4v_\Omega^2}}
\end{eqnarray}
In this case, the masses eigenvalues are
\begin{eqnarray}
m_{G^\pm}^2 &=& 0
\nonumber\\
m_{H^\pm}^2 &=& \frac{1}{2\sqrt{2}} \frac{\mu_{\phi\Omega}}{v_\Omega} \left(
v_\phi^2 + 4 v_\Omega^2 \right) \label{eq:hhiggs}
\end{eqnarray}

Here the zero mass eigenstate corresponds to the charged Goldstone
boson and the massive eigenstate corresponds to a physical charged
scalar. Since the $\rho$ parameter receives contributions from the vev
$v_\Omega$, having $\rho = 1$ constrains the vev of the triplet to be
much smaller than the Higgs-doublet vev (the limit is set to
$v_\Omega \lsim 5\,\GeV$). This implies that the Goldstone boson is
mainly doublet, while the massive charged scalar is mainly
triplet. Therefore, we can make the following approximation for
charged scalar mass,
\begin{eqnarray}
m_{H^\pm}^2 &\approx& \frac{1}{2\sqrt{2}} \frac{\mu_{\phi\Omega}}{v_\Omega} v_\phi^2 \label{eq:chhiggs}
\end{eqnarray}
One sees that by taking $\mu_{\phi\Omega} > v_\Omega$ one can have a
charged Higgs in agreement with current experiments.
In order to find the mass eigenstates for the neutral scalar mass
matrix, we follow a similar procedure. After solving the tadpoles
for $m_\phi^2$ and $m_\Omega^2$, the CP-even Higgs mass matrix in the
basis $\left(\chi_\phi , \chi_\Omega \right)^T$ becomes,
\begin{eqnarray}
M_\chi^2 &=& \Bigg[ \begin{array}{ccc} 
\lambda_\phi \, v_\phi^2 & 
\lambda_{\phi\Omega} \, v_\phi v_\Omega - \frac{1}{\sqrt{2}} \mu_{\phi\Omega} \, v_\phi \\ 
\lambda_{\phi\Omega} \, v_\phi v_\Omega - \frac{1}{\sqrt{2}} \mu_{\phi\Omega} \, v_\phi & 
2 \lambda_\Omega \, v_\Omega^2 
+ \frac{1}{2\sqrt{2}} \mu_{\phi\Omega} \, \frac{v_\phi^2}{v_\Omega}
\end{array} \Bigg]
\end{eqnarray}

In this case we have two massive eigenstates. The lightest one is
identified with the Higgs of $125\,\GeV$, while the second one will be
our heavy Higgs and the center of our following analysis.  This is
motivated by negative searches for a light Higgs boson
\cite{Aad:2014ioa,Khachatryan:2017mnf,Aad:2013yqp}.  We define
$\alpha$ as the mixing angle in the CP-even sector, which defines the
following orthogonal matrix:
%
\begin{eqnarray}
O_\chi &=& 
\left[ \begin{array}{cc} c_\alpha & - s_\alpha \\ s_\alpha & c_\alpha \end{array} \right] 
\end{eqnarray}
such that
$O_\chi M_\chi^2O_\chi^T={\text{diag}}(m_{h_1}^2,m_{h_2}^2)$, with the
mass eigenstates named $h_1$ and $h_2$ in increasing mass ordering.
Notice that we want that the first eigenstate to be
Higgs-like in order to agree with the experiments, which
force $c_\alpha$ to be larger than $s_\alpha$. In this setup, it is
immediate that the second neutral state $h_2$ will be mainly triplet.

By using again the approximation in which $v_\phi \gg v_\Omega$, and
$\mu_{\phi\Omega} > v_\Omega$, the eigenvalues of the neutral CP-even
mass matrix can be greatly simplified to:
\begin{eqnarray}
m_{h_1}^2 &\approx& \lambda_\phi v_\phi^2
\nonumber\\
m_{h_2}^2 &\approx& \frac{1}{2\sqrt{2}} \frac{\mu_{\phi\Omega}}{v_\Omega} v_\phi^2 \,
\approx \, m_{H^\pm}^2 \label{eq:nhmass}
\end{eqnarray}
So that, this setting always admits a light doublet-like Higgs ($h_1$)
and a heavy triplet-like Higgs ($h_2$), in which the heavy Higgses are
nearly degenerate, so that
$m_{h_2}^2 \approx \frac{1}{2\sqrt{2}}
\frac{\mu_{\phi\Omega}}{v_\Omega} v_\phi^2 \approx m_{H^\pm}^2$.

\section{Numerical Analysis}
\label{sec:numerical-analysis}

\subsection{Parameter Space}

We now turn to the description of the physically acceptable parameter
space of the model. 
We vary randomly the values of the different parameters, but requiring:
boundedness from below of the scalar potential, and the exact
conservation of the $\mathbb{Z}_2$ symmetry, as already been discussed
above.
Moreover we require that the lightest Higgs bosons with 125 GeV mass
is mainly doublet by demanding $|O_\chi^{11}|>0.9$ which is satisfied
by most of our generated scenarios.
With these restrictions we seek to obtain consistent neutrino physics
(Sec.\ref{sec:mnu} below) and a correct dark matter relic abundance
with a scalar dark matter candidate, Sec. \ref{sec:relic} below.
Recall that, as we stated in the end of the section \ref{sec:model},
we will pick the $\eta_R$ as the dark matter candidate of our model.

\subsubsection{Neutrino mass}
\label{sec:mnu}
 
  In our study of scotogenic dark matter we need first to
  implement the neutrino oscillation constraints. As usual, this can
  be achieved in two configurations which are called normal and
  inverse mass orderings. A feature of our singlet-triplet scotogenic
  model is that one of the neutrinos is massless, simplifying the
  discussion. For further convenience, we use the Casas-Ibarra
  parametrization~\cite{Casas:2001sr}.  Since our results do not
  depend on the detailed nature of the neutrino spectrum, we also
  choose to focus on the case of normal ordering for neutrino
  masses. This can be understood since neither dark matter nor the
  collider phenomenology are sensitive to the individual masses of
  different neutrino species, and hence to their specific mass
    hierarchy~\cite{Rocha-Moran:2016enp}.
  Small values of $\lambda''_{\phi\eta}$ were chosen by the original
  references \cite{Hirsch:2013ola,Merle:2016scw} in order to have
  small neutrino masses with ${\cal{O}}(1)$ Yukawa
  couplings~\footnote{In references
    \cite{Hirsch:2013ola,Merle:2016scw} the parameter
    $\lambda''_{\phi\eta}$ is named $\lambda_5$.}. This fits with
  t'Hooft naturalness principle, since making $\lambda''_{\phi\eta}$
  zero would deliver a Lagrangian with an extra $U(1)$ symmetry which
  is identified with lepton number. Here we allow
  $\lambda''_{\phi\eta}$ to take on larger values.  This implies the
  need for choosing small values for the Yukawa couplings so as to
  account for small neutrino masses
  \cite{Rocha-Moran:2016enp}. Moreover it implies that the degeneracy
  between $m_{\eta^R}$ and $m_{\eta^I}$ is lifted.

\begin{figure}[ht]
\begin{center}
\centerline{\protect\vbox{\epsfig{file=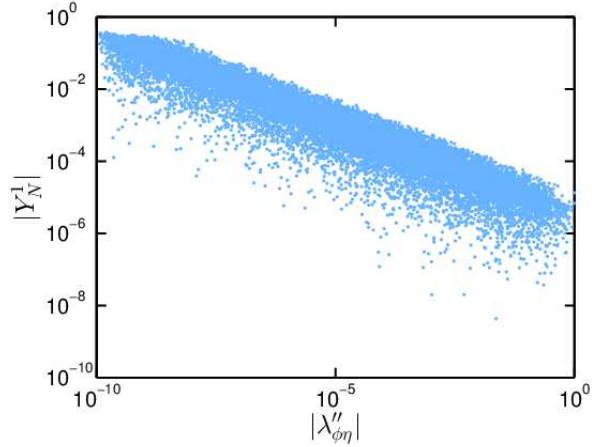,width=0.5\textwidth}}}
\caption{\it Scan where all the independent parameters of the model
  are varied. The absolute value $|Y^1_N|$ of one of the Yukawa
  couplings is presented as a function of $|\lambda''_{\phi\eta}|$.
  It shows that $|\lambda''_{\phi\eta}|$ can be large as long as the
  Yukawa couplings are small.  }
\label{fig:LambdaPPvsYn1}
\end{center}
\end{figure}
%

In Fig.~\ref{fig:LambdaPPvsYn1} we show a general scatter plot (all
model parameters are randomly varied) for points in parameter space
that satisfy the measured neutrino oscillation parameters and other
constraints. It shows the Yukawa coupling $|Y^1_N|$ as a function of
$|\lambda''_{\phi\eta}|$.  One sees that a large Yukawa coupling
requires a small value for $|\lambda''_{\phi\eta}|$ and vice-versa. A
similar behavior is observed in the other Yukawa couplings, in fact,
since neutrino masses have well defined values, there is a correlation
between the Yukawas (encoded in the matrix $h$) and
$\lambda_{\phi\eta}''$.  In the limit where
$\lambda_{\phi\eta}''v_{\phi}^{2} \ll m_0^2,\, m_{\zeta_\sigma}^2$ and
$\lambda_{\phi\eta}''v_{\phi}^{2} \ll m_0^2\,-\, m_{\zeta_\sigma}^2$,
we find that
\begin{equation}
{\cal M}_\nu^{\alpha\beta} = \sum_{\sigma=1}^2 \frac{h^{\alpha\sigma}h^{\beta\sigma}}{32\pi^2}
m_{\zeta_\sigma} \frac{\lambda_{\phi\eta}'' v_\phi^2}{m_{\zeta_\sigma}^2\, -\,m_0^2} \mathcal{F}\left( m_{\zeta_\sigma}, m_0 \right) \label{eq:corrYL}
\end{equation}
Where the function $\mathcal{F}$ involves only logarithms of the quantities $m_{\zeta_\sigma}$ and $m_0$,
and in turn (c.f. Eqs.~(\ref{eq:etaR}) and Eqs.~(\ref{eq:etaI})):
\begin{equation}
m_{0}^{2} \equiv m_\eta^2 + 
\frac{1}{2} (\lambda_{\phi\eta}+\lambda'_{\phi\eta}) v_\phi^2
+ \frac{1}{2} \lambda_{\eta\Omega} \, v_\Omega^2
- \frac{1}{\sqrt{2}} \mu_{\eta\Omega} \, v_\Omega
\end{equation}
Actually, the correlation between $Y_{N,\Sigma}$ and
$\lambda_{\phi\eta}''$ is general and shows up basically in any limits
that one may take in eq.~(\ref{eq:mnuscoto}) \cite{Hirsch:2012ym}.

\subsubsection{Relic density}
\label{sec:relic}

Due to the presence of the $Z_2$ symmetry we have three possible dark matter
candidates in our model: $\eta_R$, $\eta_I$, see
eq.~(\ref{eq:scalars})) or $\zeta_1$, see eq.~(\ref{eq:fermions}).
The lightest of these states is a potential WIMP DM candidate within
our scotogenic scenario.
For example, the possibility of the fermion $\zeta_1$ being the dark
matter candidate has been studied before~\cite{Hirsch:2013ola}.
For definiteness, in our following analysis we assume the alternative
possibility that the DM is the $\eta_R$ field and study its viability.

\begin{figure}[ht]
\begin{center}
\centerline{\protect\vbox{
\epsfig{file=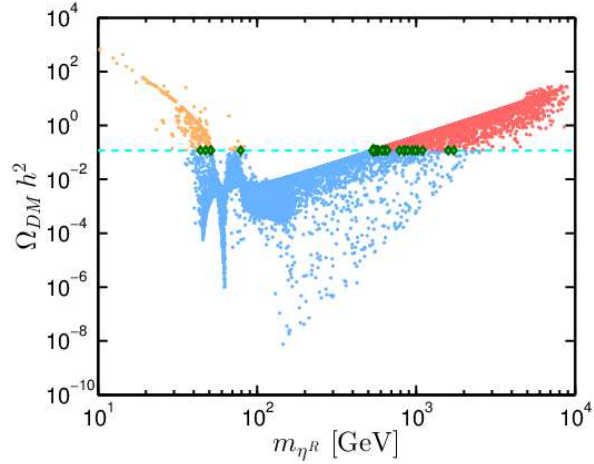,width=0.5\textwidth}}
}
\caption{\it Relic density for our dark matter candidate $\eta_R$ as a function
  of its mass.  We show a scan varying all the independent parameters
  of the model. Only points with $h_1$ mainly doublet are
  displayed. See the text for the color code.  }
\label{fig:Omega_etR}
\end{center}
\end{figure}
%
In Fig.~\ref{fig:Omega_etR} we show a plot of the dark matter relic
abundance as a function of the mass of the scalar dark matter
candidate, which is taken to be $\eta_R$. Only points obeying
$|c_\alpha|\equiv|O_\chi^{11}|>0.9$ are shown, because experimental
results seem to prefer a SM-like $125\,\GeV$ Higgs boson
\cite{Aad:2015zhl}.  The measured value for the relic abundance lies
in a small region which is shown as a cyan horizontal line in the
figure \cite{Bennett:2012zja,Ade:2015xua}.  The color code for the
points is explained as follows.  Points that fulfill the measured
value of the relic density are painted in green.  In blue, we have
points that have a relic abundance under the $3\sigma$ region on the
current measurements; the red and orange ones have a relic abundance
over the mentioned region. Nevertheless, the
blue points are only ruled out if we consider the $\eta_R$ as the only
source of Dark Matter.
Notice that this model has similar results for direct dark matter
detection to those of the IHDM model. In fact, one sees two dips in the
relic abundance plot in fig.~\ref{fig:Omega_etR}, which correspond to
the situations where the DM has resonant scattering through a Higgs or
Z boson at $\sim 45\,\GeV$ and $\sim 62.5\,\GeV$. The same features
are also present in IHDM~case~\cite{Diaz:2015pyv}.
Concerning the phenomenology of dark matter in this model it is worth
mentioning that scalar dark matter can be detected directly by nuclear
recoil through the Higgs portal mechanism, as well as at the LHC and
indirectly~\cite{Djouadi:2012zc,Hirsch:2012ym,Chulia:2016giq,Diaz:2015pyv,Bonilla:2016diq}.
For instance the mass gap that separates red and orange points in the
present model has its analogous gap in that case too. This is not
surprising since both models share many features. 
%
\begin{figure}[ht]
\bigskip
\centering
\unitlength = 1mm
\begin{fmffile}{Diagrams/scalar_AC_1}
\fmfset{arrow_len}{0mm}
\begin{fmfgraph*}(25,30)
\fmfleft{i1,i2}
\fmfright{o1,o2}
\fmflabel{$\eta_{R,I}$}{i1}
\fmflabel{$\eta_R$}{i2}
\fmflabel{$\nu$}{o1}
\fmflabel{$\nu$}{o2}
\fmf{dashes}{i1,v1}
\fmf{dashes}{i2,v2}
\fmf{fermion}{o1,v1}
\fmf{fermion,label=$\zeta_{\beta}$}{v1,v2}
\fmf{fermion}{v2,o2}
\end{fmfgraph*}
\end{fmffile}
\hspace{1cm}
\unitlength = 1mm
\begin{fmffile}{Diagrams/scalar_AC_2}
\fmfset{arrow_len}{0mm}
\begin{fmfgraph*}(25,30)
\fmfleft{i1,i2}
\fmfright{o1,o2}
\fmflabel{$\eta^{\pm}$}{i1}
\fmflabel{$\eta_R$}{i2}
\fmflabel{$e^{\pm}$}{o1}
\fmflabel{$\nu$}{o2}
\fmf{dashes}{i1,v1}
\fmf{dashes}{i2,v2}
\fmf{fermion}{o1,v1}
\fmf{fermion,label=$\zeta_{\beta}$}{v1,v2}
\fmf{fermion}{v2,o2}
\end{fmfgraph*}
\end{fmffile}
\hspace{1cm}
\unitlength = 1mm
\begin{fmffile}{Diagrams/scalar_AC_3}
\fmfset{arrow_len}{0mm}
\begin{fmfgraph*}(25,30)
\fmfleft{i1,i2}
\fmfright{o1,o2}
\fmflabel{$\eta_{R,I}$}{i1}
\fmflabel{$\eta_R$}{i2}
\fmflabel{$e^{\pm}$}{o1}
\fmflabel{$e^{\mp}$}{o2}
\fmf{dashes}{i1,v1}
\fmf{dashes}{i2,v2}
\fmf{fermion}{o1,v1}
\fmf{fermion,label=$\zeta^{\pm}$}{v1,v2}
\fmf{fermion}{v2,o2}
\end{fmfgraph*}
\end{fmffile}
\hspace{1cm}
\unitlength = 1mm
\begin{fmffile}{Diagrams/scalar_AC_4}
\fmfset{arrow_len}{0mm}
\begin{fmfgraph*}(25,30)
\fmfleft{i1,i2}
\fmfright{o1,o2}
\fmflabel{$\eta^{\pm}$}{i1}
\fmflabel{$\eta_R$}{i2}
\fmflabel{$\nu$}{o1}
\fmflabel{$e^{\pm}$}{o2}
\fmf{dashes}{i1,v1}
\fmf{dashes}{i2,v2}
\fmf{fermion}{o1,v1}
\fmf{fermion,label=$\zeta^{\mp}$}{v1,v2}
\fmf{fermion}{v2,o2}
\end{fmfgraph*}
\end{fmffile}
\bigskip
\caption{llustrative Feynman diagrams for scalar
  annihilation/coannihilation channels involving Yukawa couplings
  $Y_N$ and $Y_\Sigma$.}
\label{fig:co-annihilation}
\end{figure}
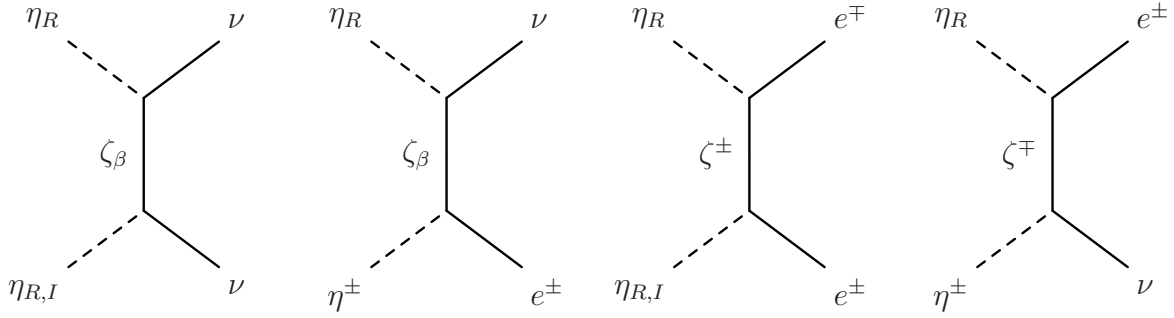
%
\begin{figure}[ht]
\begin{center}
\centerline{\protect\vbox{\epsfig{file=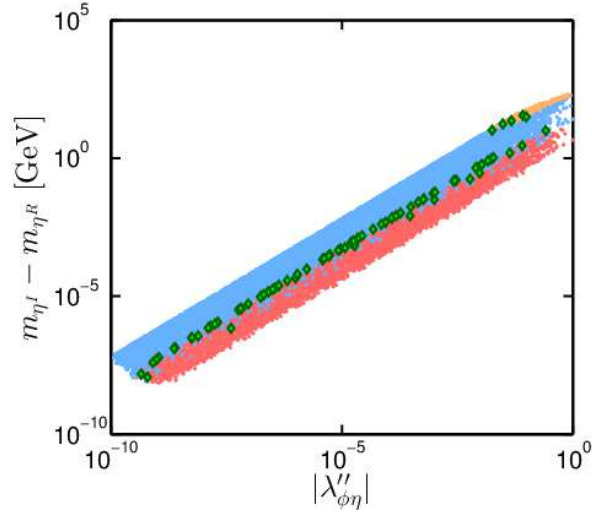,width=0.5\textwidth}
}}
\caption{\it Mass difference $m_{\eta^I}-m_{\eta^R}$ as a function of
  $|\lambda''_{\phi\eta}|$ with identical color code as in the
  previous figure.  Interesting is to notice that low values of
  $m_{\eta_R}$ with an excess of relic density are associated to high
  values of $|\lambda''_{\phi\eta}|$.  }
\label{LambdaPPvsDeltaMwRD}
\end{center}
\end{figure}
%
The difference $m_{\eta^I} - m_{\eta^R}$ is intimately connected with
the violation of lepton number and the value of
$\lambda''_{\phi\eta}$, as it can be seen from equations
(\ref{eq:etaR}) and (\ref{eq:etaI}). In fact, by taking
  the limit of small $\lambda_{\phi\eta}''$ in these equations one
  finds
\begin{equation}
m_{\eta^{R}}\, -\,m_{\eta^{I}} \approx \frac{2\lambda_{\phi\eta}'' v_{\phi}^{2}}{\sqrt{3}m_{0}}. \label{eq:deg}
\end{equation}
This implies a relation between dark matter relic abundance and
neutrino physics, since the Yukawas $Y_{\Sigma}$ and $Y_{N}$ are
involved both in the neutrino mass generation as in the DM
annihilation cross section. For instance, the annihilation
  $\eta_R \eta_R \to \nu\nu$ can be seen from the Lagrangian
  \ref{eq:yukawa}, in which the Yukawa couplings $Y_N$ and $Y_\Sigma$
  indicate interactions among the DM candidate, neutrinos and the
  neutral fermions $\Sigma_0$ and $N$, so that they lead to a
  t-channel contribution to the annihilation cross section. Scalar
  annihilation/coannihilation channels generated by the Yukawa
  couplings $Y_N$ and $Y_\Sigma$ are shown in
  Fig.~\ref{fig:co-annihilation}.\\

In Fig.~\ref{LambdaPPvsDeltaMwRD} we see
the mass difference $m_{\eta^I}-m_{\eta^R}$ as a function of
$|\lambda_{\phi\eta}''|$, with the same color code as in the previous
figure. We notice that the orange points, associated to large values
of $|\lambda_{\phi\eta}''|$ in Fig.~\ref{LambdaPPvsDeltaMwRD} are also
associated to the small values of the dark matter mass in
Fig.~\ref{fig:Omega_etR}.  Therefore, small values for $m_{\eta^R}$
(DM) need very small Yukawa couplings, $Y_N$ and $Y_\Sigma$, reason
why we consider them disfavored. Notice also that many points that
give the correct relic density lie in the whole range for
$|\lambda_{\phi\eta}''|$, approximately at the intersection of the
blue and red regions.
Notice also that eq.(\ref{eq:deg}) shows that there are two ways to
enforce the degeneracy between $\eta_R$ and $\eta_I$ fields. One of
them is the smallness of $|\lambda_{\phi\eta}''|$, and the other the
largeness of $m_0$. This degeneracy is important because it determines
the smallness of neutrino mass. Unfortunately, however, it is not
directly translated into a prediction for the relic abundance. Indeed,
all the blue and red points correspond to degeneracy, while the
orange ones do not.

\subsection{Production of heavy Higgs boson}

\subsubsection{ Two benchmark points}

\begin{figure}[ht]
\begin{center}
\centerline{\protect\vbox{\epsfig{file=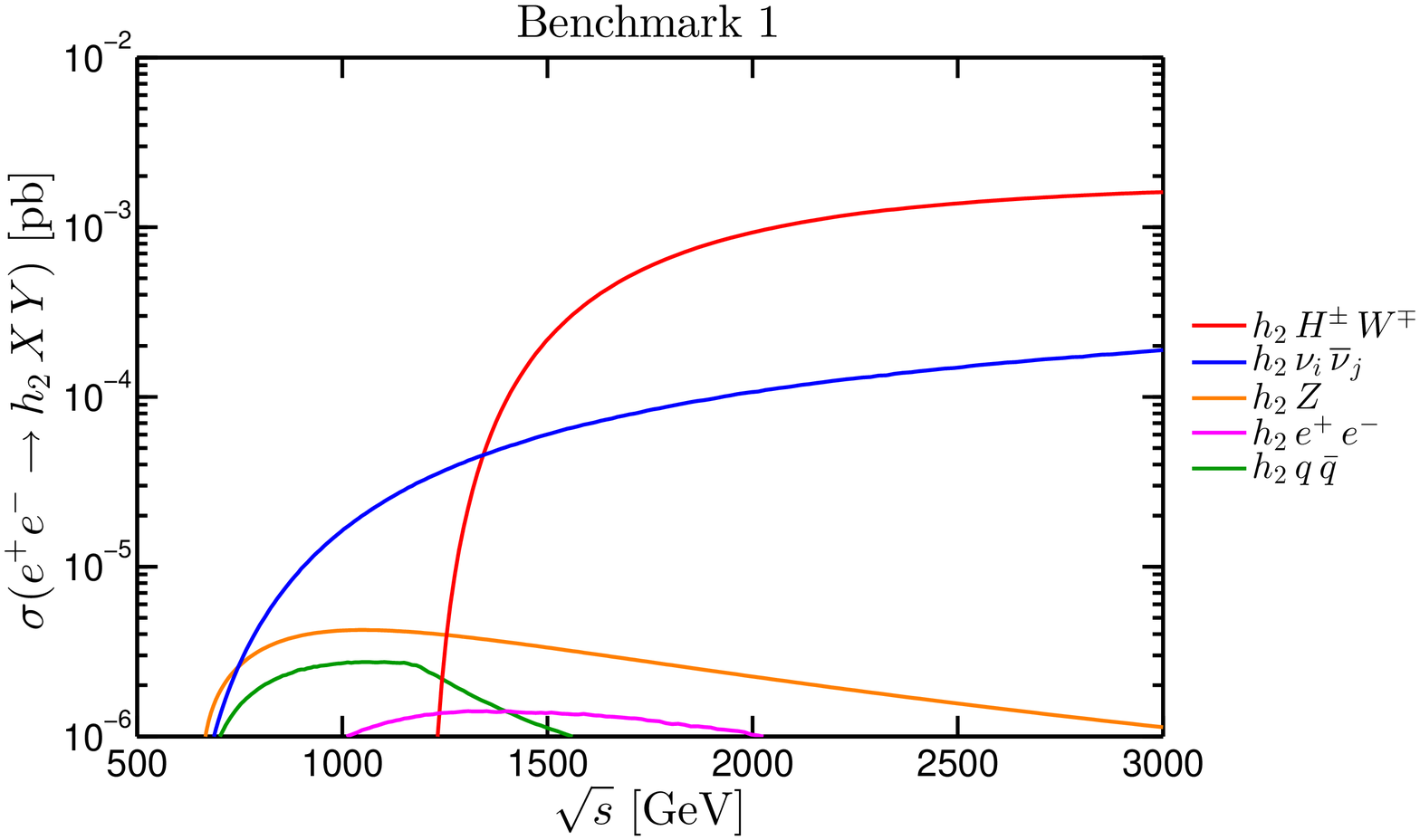,width=0.45\textwidth}
\epsfig{file=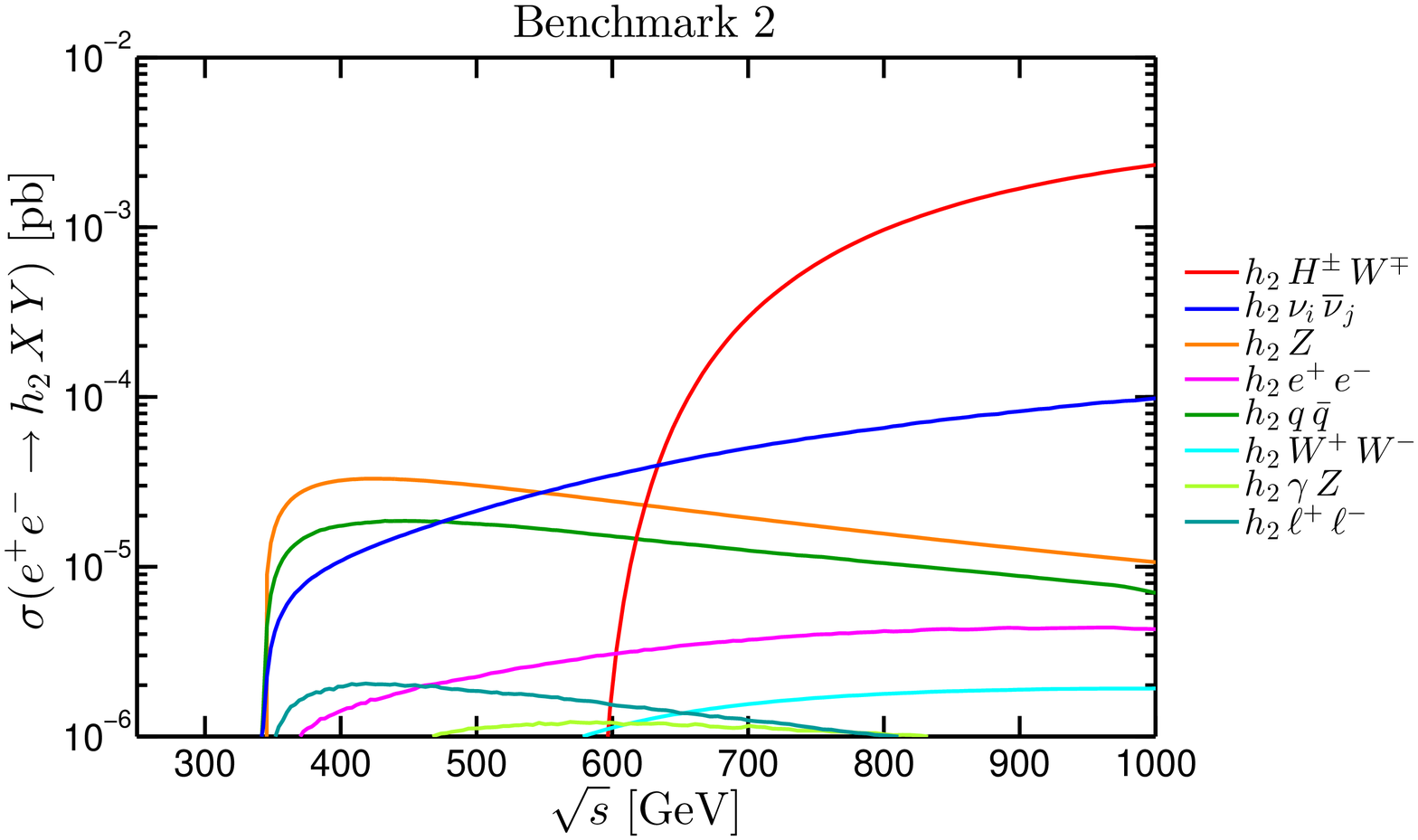,width=0.45\textwidth}}}
\caption{\it Two and three body production cross section for $h_2$ in
  an electron-positron collider as a function of $\sqrt{s}$ for both
  benchmarks.  }
\label{fig:CS_h2xy}
\end{center}
\end{figure}

In this section we study the phenomenology of the heavy neutral Higgs
boson $h_2$ present in this model. We choose two different benchmarks,
that we call B1 and B2. They are required to be consistent with a
scalar DM candidate and the other constraints described above.  B1 is
given in Table \ref{tab:tab1} and the corresponding scalar spectrum is
given in Table~\ref{tab:tab2}.
In order to obtain a lighter Higgs boson spectrum, so as to study the
branching ratios and LHC production cross sections, we define B2 by
starting from B1 and changing only one parameter. 
This parameter is chosen to be $\mu_{\phi\Omega}$, since it controls
the masses of $H^{+}$ and $h_{2}$, as shown in eqs.~(\ref{eq:chhiggs})
and (\ref{eq:nhmass}).  
We vary it from $\mu_{\phi\Omega}=54$ GeV in B1 to
$\mu_{\phi\Omega}=11$ GeV in B2. This change produces a lighter
spectrum in B2 as compared to B1, as seen in
Table~\ref{tab:tab2}. Other possible benchmark points hardly affect
the cross sections of interest to us,
 except that of the two largest
 cross sections (see fig.~\ref{fig:CS_h2xy}),
$\sigma\left( e^+ e^- \longrightarrow h_2 H^{\pm}W^{\mp} \right)$
rises with the parameter $v_\Omega$, while
$\sigma\left( e^+ e^- \longrightarrow h_2 \nu_i \overline{\nu}_j
\right)$
is  unaffected. Therefore, since we know the effect of varying
$v_\Omega$, we did not include a benchmark with a smaller $v_\Omega$.

The neutrino sector is unchanged, and the same Yukawa couplings
are chosen in both benchmarks in order to satisfy neutrino mass square
differences and mixing angles required in order to account for
neutrino oscillation data~\cite{Forero:2014bxa}.  The Yukawa couplings
are also given in the table.
\begin{table}[!h]
\begin{center}
\begin{tabular*}{0.6\textwidth}{@{\extracolsep{\fill}} c c c c}
\hline\hline
Parameter              & Benchmark 1 & Benchmark 2 & Units \\ \hline
$v_\phi$               & $246$       & $246$       & GeV   \\
$v_\Omega$             & $3.7$       & $3.7$       & GeV   \\
$\lambda_\phi$         & $0.263$     & $0.263$     & -     \\
$\lambda_\eta$         & $0.41$      & $0.41$      & -     \\
$\lambda_\Omega$       & $0.65$      & $0.65$      & -     \\
$\lambda_{\phi\eta}$   & $0.75$      & $0.75$      & -     \\
$\lambda_{\phi\eta}'$  & $-0.86$     & $-0.86$     & -     \\
$\lambda_{\phi\eta}''$ & $-0.0041$   & $-0.0041$   & -     \\
$\lambda_{\phi\Omega}$ & $0.47$      & $0.47$      & -     \\
$\lambda_{\eta\Omega}$ & $0.82$      & $0.82$      & -     \\
$\mu_{\phi\Omega}$     & $54$        & $11$        & GeV   \\
$\mu_{\eta\Omega}$     & $910$       & $910$       & GeV   \\
$|m_\eta|$             & $1690$      & $1690$      & GeV   \\
\hline\hline
$Y_\Omega$   & $-2.7\times10^{-1}$   & $-2.7\times10^{-1}$  & - \\
$Y_N^1$      & $1.1\times10^{-4}$    & $1.1\times10^{-4}$   & - \\
$Y_N^2$      & $1.0\times10^{-4}$    & $1.0\times10^{-4}$   & - \\
$Y_N^3$      & $-1.3\times10^{-4}$   & $-1.3\times10^{-4}$  & - \\
$Y_\Sigma^1$ & $-1.0\times10^{-4}$   & $-1.0\times10^{-4}$  & - \\
$Y_\Sigma^2$ & $-5.0\times10^{-4}$   & $-5.0\times10^{-4}$  & - \\
$Y_\Sigma^3$ & $-4.4\times10^{-4}$   & $-4.4\times10^{-4}$  & - \\
\hline\hline
\end{tabular*}
\caption{Independent parameters for the benchmark, relevant for the
  scalar sector, including dark matter (above), and the Yukawa sector
  (below).
  \label{tab:tab1}}
\end{center}
\end{table}
The scalar spectra in both benchmarks is given in table
\ref{tab:tab2}.  The particles $h_2$ and $H^+$ are both mainly
triplet, and as a result their masses are similar irrespective of the
benchmark (in B1 $H^+$ is 6 MeV lighter than $h_2$).  As mentioned
before, $\eta_R$ and $\eta_I$ have a very similar mass too, and this
is related to the symmetry that dictates the smallness of the neutrino
masses. 
\begin{table}[!h]
\begin{center}
\begin{tabular*}{0.5\textwidth}{@{\extracolsep{\fill}} c c c}
\hline\hline
Parameter    & Benchmark 1 & Benchmark 2 \\ \hline
$m_{h_1}$    & $125$  & $125$  \\
$m_{h_2}$    & $560$  & $253$  \\
$m_{H^+}$    & $560$  & $252$  \\
$m_{\eta^R}$ & $1688$ & $1688$ \\
$m_{\eta^I}$ & $1688$ & $1688$ \\
$m_{\eta^+}$ & $1697$ & $1697$ \\
\hline\hline
\end{tabular*}
\caption{Scalar physical masses in GeV for the chosen benchmarks.
\label{tab:tab2}}
\end{center}
\end{table}

In both benchmarks $\eta_I$ is around 80 MeV heavier than
$\eta_R$ (our dark matter particle). Thus, the largest difference
between the two benchmarks is that the new scalars are relatively
heavier in B1.
%

Using the latest version of Madgraph \cite{Alwall:2014hca} we
calculate the $h_2$ production cross sections for both benchmarks. The
results are displayed in Fig.~\ref{fig:CS_h2xy}. There we give 2-body
and 3-body production cross sections as a function of the center of
mass energy (c.m.) $\sqrt{s}$ in a electron-positron collider. To
evaluate the reach of a couple of planned electron-positron colliders
we mention that the projected c.m. energies and luminosities for the
International Linear Collider (ILC) are 250 GeV, 500 GeV, and 1000
GeV, with 250 fb$^{-1}$, 500 fb$^{-1}$, and 1000 fb$^{-1}$
respectively \cite{Baer:2013cma}.
For the case of the Compact Linear Collider (CLIC) these are
$\sqrt{s}=350$ GeV, 1.4 TeV, and 3 TeV, with estimated luminosities
$500 {\,\mathrm{fb}}^{-1}$, $1.5. {\, \mathrm{ab}}^{-1}$, and
$2 {\, \mathrm{ab}}^{-1}$ respectively \cite{Abramowicz:2013tzc}.

We see in Fig.~\ref{fig:CS_h2xy} that the only relevant 2-body $h_2$
production mechanism is $e^+\,e^- \to h_2\,Z$, which is suppressed up
to 3 orders of magnitude at $\sqrt{s}=3$ TeV for B1, and suppressed up
to 2 orders of magnitude at $\sqrt{s}=1$ TeV for B2, compared to the
main 3-body production mode at those energies.  The channel
$e^+\,e^- \to h_2\,Z$ can dominate at low energies though, as can be
seen in the B2 frame. Concerning 3-body production modes, the largest
two are $e^+\,e^-\to h_2 H^\pm W^\mp$ (including a factor of 2 due to
the choice of electric charge) and $e^+\,e^-\to h_2 \nu_i \bar\nu_j$
in both benchmarks.  The production process $e^+\,e^-\to h_2 H^+ W^-$
is in turn dominated by the sub-process where a neutrino is in the
t-channel and a $W$ boson produces the pair $h_2 H^+$. Thus, the
coupling $h_2 H^+ W^-$ becomes important. Since $h_2$ and $H^+$ are
mainly triplet, this coupling is not suppressed with respect to the
gauge coupling.  The 3-body production mode
$e^+\,e^-\to h_2 \nu_i \bar\nu_j$ is one of the two known as ``Vector
Boson Fusion'' (VBF), because the mentioned process is dominated by a
sub-process where a $W$ boson is emitted from the electron (and
positron) and they ``fuse'' to create an $h_2$.  If we replace the $W$
boson by a $Z$ boson we find the second VBF process
$e^+\,e^-\to h_2 e^+ e^-$. It is worth mentioning that this last VBF
is suppressed with respect to $e^+\,e^-\to h_2 \nu_i \bar\nu_j$, as
can be seen in Fig.~\ref{fig:CS_h2xy}, because most of the charged
leptons go through the beam pipe.  If we use the corresponding cut
(Madgraph has this cut by default), the cross section diminishes
considerably \cite{Blunier:2016peh}. As with the 2-body production
mode, $e^+\,e^-\to h_2 \nu_i \bar\nu_j$ can be the dominant process at
low energies.  We also note the reader that we have checked that when
we decrease the triplet vev $v_\Omega$ the cross section
$\sigma(e^+\,e^-\to h_2 H^+ W^-)$ is hardly affected while the cross
section $\sigma(e^+\,e^-\to h_2 \nu_i \bar\nu_j)$
decreases, as expected from the couplings involved.\\

If we naively compare the production cross sections shown in
Fig.~\ref{fig:CS_h2xy} with the projected luminosities of ILC and CLIC
given in the immediately following paragraph we conclude that ILC
will have the chance to observe $h_2$ in this model only if its mass
is low (as given in B2).  On the contrary, CLIC will have enough
energy and luminosity to observe $h_2$ in both benchmarks
\cite{Blunier:2016peh}.

%
\begin{figure}[ht]
\begin{center}
\centerline{\protect\vbox{\epsfig{file=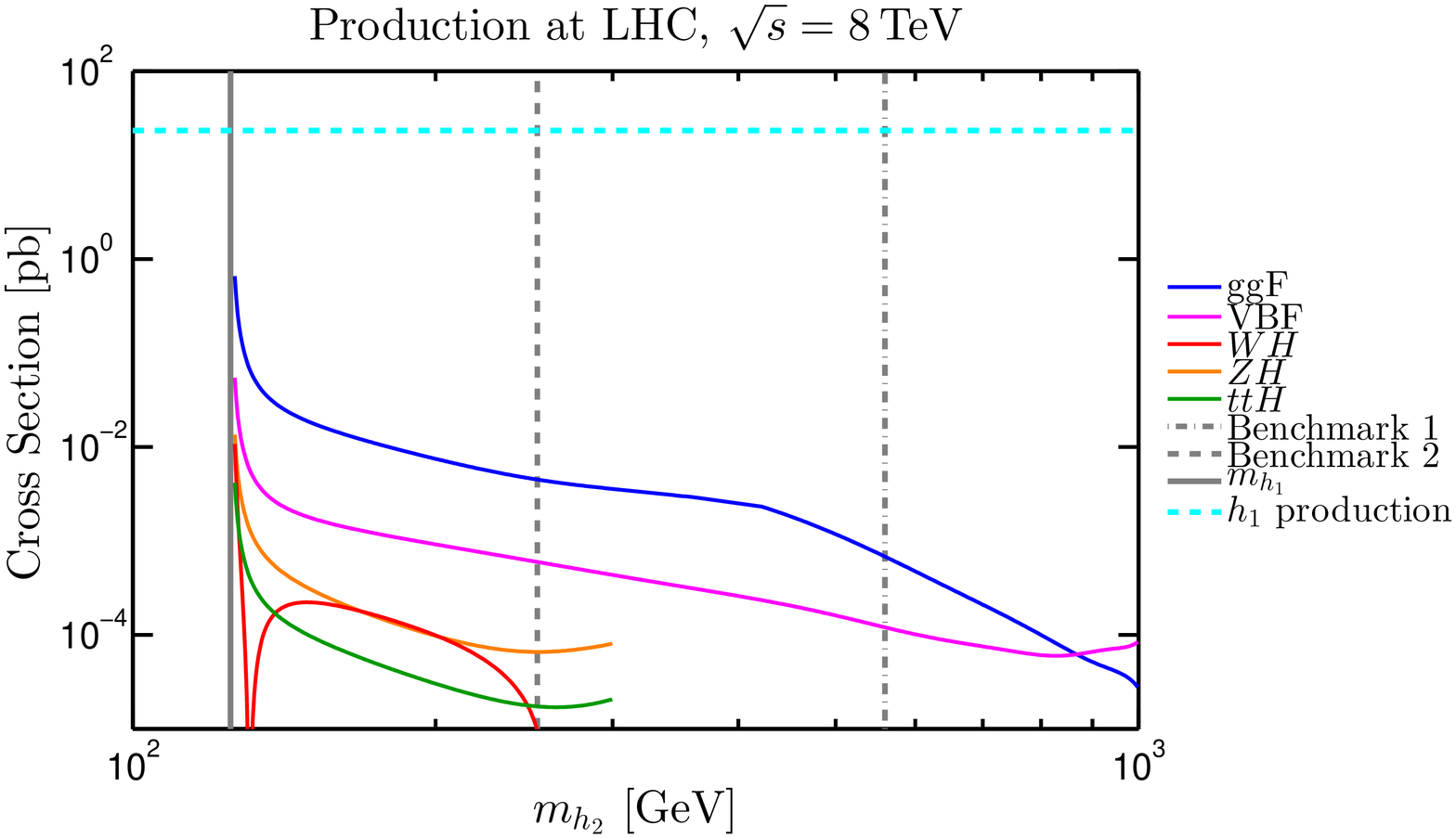,width=0.45\textwidth}
\epsfig{file=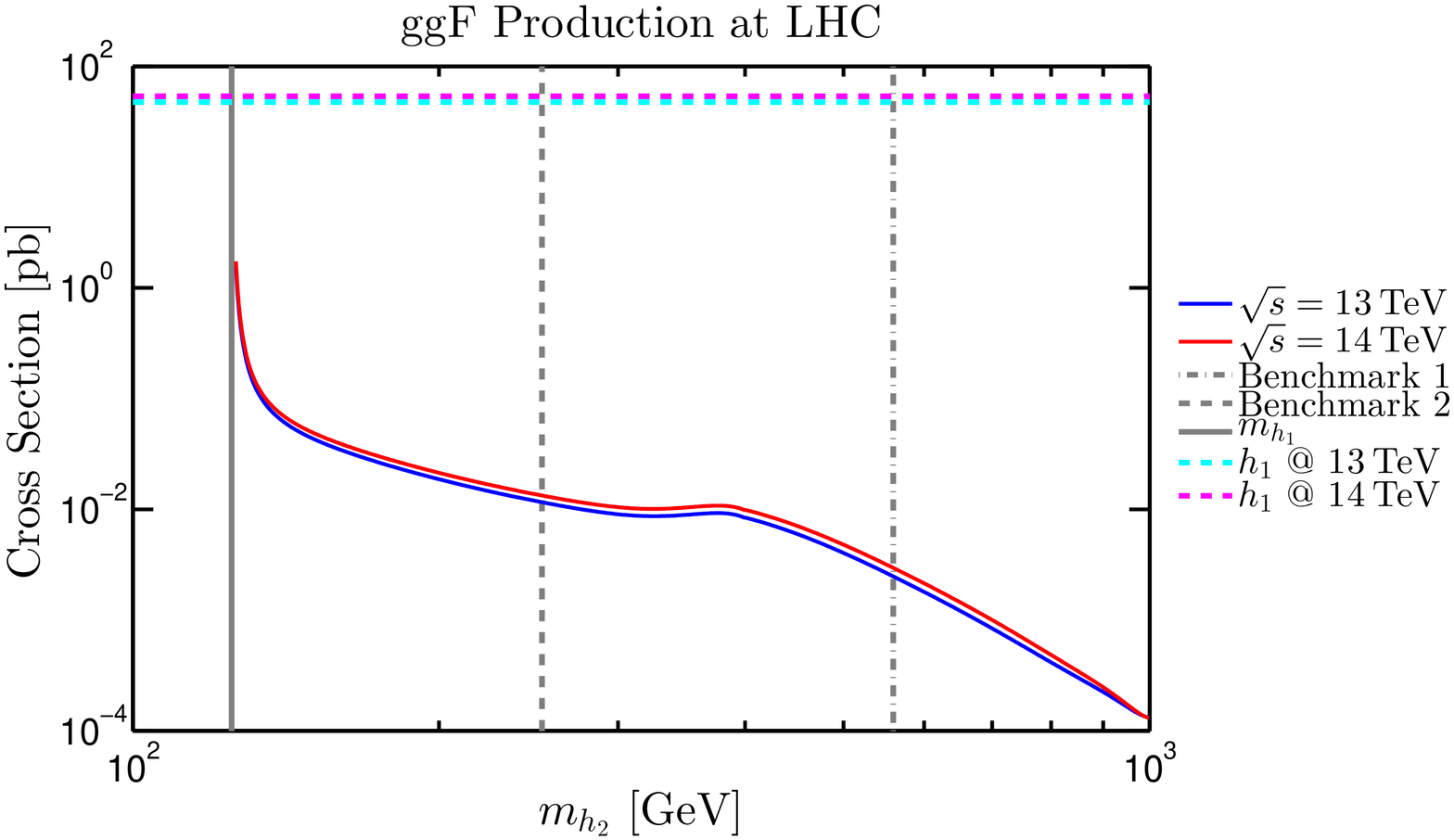,width=0.45\textwidth}}
}
\caption{\it Cross section at the LHC. The vertical grey dashed lines correspond
to the value of $m_{h_2}$ in our benchmarks. The left panel
shows the cross sections computed for $\sqrt{s}=8\,\TeV$ and the right
panel is for $\sqrt{s}=13$ and $14\,\TeV$.
}
\label{fig:CS_lhc_h2}
\end{center}
\end{figure}
%
We argue that it will be very difficult to observe $h_2$ at the
LHC. In Fig.~\ref{fig:CS_lhc_h2} we see the production cross section
at the LHC for $h_2$ as a function of its mass. In the left frame we
have it for the center of mass energy $\sqrt{s}=8$ TeV, and in the
right frame we have it for 13 and 14 TeV c.m. energy. In both cases we
have the gluon-gluon fusion (ggF) production mode, while for the low
energy (8 TeV) we have also other less important production modes. The
two vertical dashed lines indicate the value of $m_{h_2}$ in both
benchmarks, and for reference we also show in a vertical solid line
the value of the SM-like Higgs 125 GeV. The horizontal dashed lines
indicates the value of the LHC production cross section of a 125 GeV
SM-like Higgs boson.

The main message of Fig.~\ref{fig:CS_lhc_h2} is that independently of
the c.m. energy, the production cross section of $h_2$ at the LHC is 3
to 5 orders of magnitude (depending on $m_{h_2}$) smaller than the
production cross section of the SM-like Higgs boson. The principal
reason for this behaviour is that the ggF production mode is based on
the $h_2$ coupling to quarks. Since only the Higgs doublet couple to
fermions, and since $h_2$ is mainly triplet, that coupling is highly
suppressed. Therefore, any hadron collider that relies on ggF for the
production of the scalar will have a hard time to observe $h_2$.

\subsubsection{Decay Rates for the Charged and Heavy CP-Even Higgs}

In this section we calculate the decay rates for $h_2$ and $H^+$ using
the latest version of SPheno code \cite{Porod:2003um}, where our model
is implemented.
%
\begin{figure}[ht]
\begin{center}
\centerline{\protect\vbox{\epsfig{file=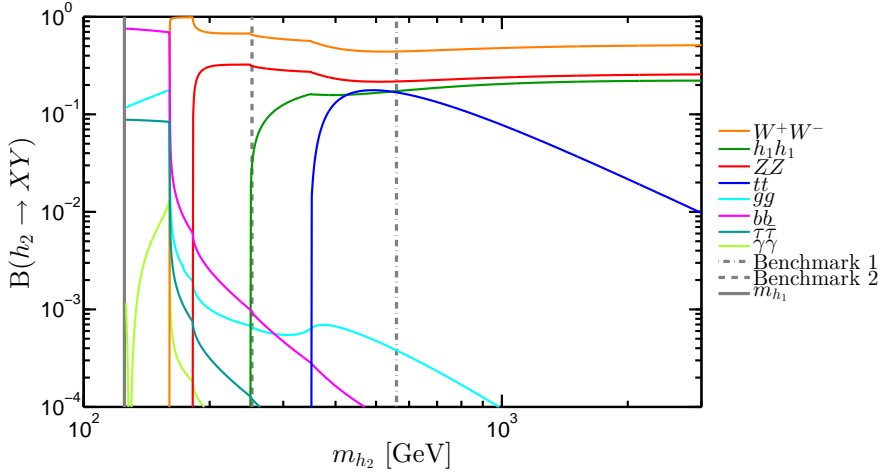,width=0.75\textwidth}
}}
\caption{\it Branching ratios for $h_2$ as a function of its mass. When the 
channels are open, the main decay modes are $h_2\to WW$ and $h_2\to ZZ$.}
\label{fig:Br_h2}
\end{center}
\end{figure}
%
In Fig.~\ref{fig:Br_h2} we have the 2-body decay modes for $h_2$ as a
function of its mass.  The solid, dashed, and dot-dashed vertical
lines indicate the 125 GeV mass for the SM-like Higgs boson, and the
value for $m_{h_2}$ in our two benchmarks. We create each branching
ratio line varying only one parameter, $\mu_{\phi\Omega}$.  Thus, at
the intersection with the corresponding vertical line, we sit exactly
at each of the two benchmarks.

Above the $ZZ$ threshold, the dominant decay mode is
$h_2\to WW$ followed by $h_2\to ZZ$. The first decay
follows because the Higgs $h_2$ is mostly triplet and has interactions
with two $W$ bosons via the Higgs kinetic term in the Lagrangian. The
decay rate for $h_2\to ZZ$ is penalized because the triplet
has null hypercharge and only couples to two $Z$ bosons through the
$h_2$ component to doublet. The third important decay mode is
$h_2\to h_1h_1$. This decay mode is important because we have
chosen a large value for $\mu_{\phi\Omega}$. In addition,
$\lambda_{\phi\Omega}$ is large, although its contribution is
proportional to the (small) triplet vev $v_\Omega$. The enhancement of
the decay $h_2\to \bar{t}t$, which also proceeds via the $h_2$
component to doublet, and is given by the contribution of (large)
Yukawa coupling. Finally, the similar behavior of the channels
involving gauge bosons in the final state and the channel with two
light Higgs relies on a proportionality of $m_{h_2}^3$ of the three
channels, being the coupling $O_{h_2 h_1 h_1}$ proportional to
$\mu_{\phi\Omega}$ which in turn is proportional to $m_{h_2}$ (see 
eqs. (\ref{eq:h2WW}) and (\ref{eq:h2h1h1})).

%
\begin{figure}[ht]
\begin{center}
\centerline{\protect\vbox{\epsfig{file=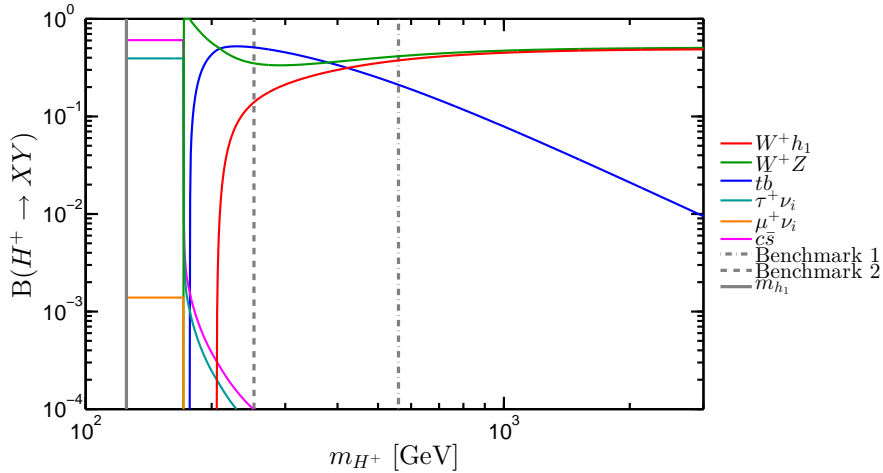,width=0.75\textwidth}
}}
\caption{\it Branching ratios for $H^+$ as a function of its mass. When the 
channels are open, the main decay modes are $H^+\to W^+Z$, $H^+\to t\bar{b}$, and 
$H^+\to W^+h_1$. }
\label{fig:Br_Hpm}
\end{center}
\end{figure}
%
In Fig.~\ref{fig:Br_Hpm} we show the decay channels for the charged
Higgs boson as a function of its mass following the same technique
described earlier for the previous case.  We are interested in the
decay modes of the charged Higgs because production cross sections for
$h_2$ are enhanced when it is accompanied by a charged Higgs, as can
be appreciated in Fig.~\ref{fig:CS_h2xy}. Above the $tb$ threshold,
three decay modes take turns in being the dominant one:
$H^+\to ZW^+$, $H^+\to t\bar{b}$, and
$H^+\to h_1 W^+$.  From the coupling point of view, all three
decay rates are small, but this is not necessarily reflected in
branching ratios. the coupling $H^+ZW^-$ is proportional to the
triplet vev, the coupling $H^+t\bar{b}$ is proportional to the $H^+$
component to doublet, and the coupling $H^+h_1W^-$ is small because
$H^+$ is mostly triplet while $h_1$ is mostly doublet.  In the same
way as the channels for a heavy neutral Higgs, the decay modes of the
charged Higgs going to $W^+ h_1$ and $W^+ Z$ behave as $m_{H^+}^3$
at large masses of this field, as it can be concluded from eqs
(\ref{eq:HpWh1}) and (\ref{eq:GmHpWZ}).

\section{Conclusions}
\label{sec:conclusions}

We have considered the proposal that the dark matter particle is a
scalar WIMP messenger associated to neutrino mass generation, whose
stability follows from the same symmetry responsible for the radiative
origin of neutrino mass.
This hypothesis embodies a simple model for WIMP dark matter and
provides a useful benchmark for electroweak symmetry breaking studies
at accelerators.
This picture was illustrated within the singlet-triplet scotogenic
dark matter model.
We have studied in detail the symmetry breaking sector of this model
and the corresponding pattern of WIMP interactions, showing how it can
provide adequate neutrino masses and dark matter relic density.
We have characterized the expected profile of heavy Higgs boson
physics as expected at the LHC and at future linear Colliders.
Our study constitutes a first step towards a comprehensive approach to
the idea that WIMP dark matter emerges as a messenger connected to
neutrino mass generation, which should encourage dedicated simulations
of the associated signatures.

\begin{center}
\section*{Acknowledgments}
\end{center}
  This work was partly funded by the Spanish grants FPA2014-58183-P,
  Multidark CSD2009-00064, SEV-2014-0398 (MINECO) and
  PROMETEOII/2014/084 (Generalitat Valenciana). MAD was partly funded by the 
  Fondecyt Grant 1141190. NR was funded by becas de postdoctorado en el 
  extranjero Conicyt/Becas Chile (2015) 74150028.\\

\appendix

\section{Appendix : Feynman Rules for Decay Channels}

{}In this section we list some of the Feynman rules of the model
relevant to evaluate the decays used in this paper.  For generality
intergenerational mixing was included, though not strictly necessary
for this paper. Throughout these diagrams, all the momenta are
incoming.

\subsection{Diagrams Involving Neutral Higgses}
\begin{eqnarray} 
\parbox{40mm}{ 
\begin{fmffile}{Diagrams/FeynDia58} 
\fmfframe(20,20)(20,20){ 
\begin{fmfgraph*}(75,75) 
\fmfleft{l1}
\fmfright{r1,r2}
\fmf{dashes}{l1,v1}
\fmf{boson}{v1,r1}
\fmf{boson}{r2,v1}
\fmflabel{$h_{{i}}$}{l1}
\fmflabel{$W^-_{{\sigma}}$}{r1}
\fmflabel{$W^+_{{\mu}}$}{r2}
\end{fmfgraph*}} 
\end{fmffile}}
&=& iO_{h_i W W} g_{\sigma \mu} \nonumber \\
&=& \frac{i}{2}\, g_{\sigma \mu}\, g_{2}^{2}\, \left(4 v_\Omega\, O_{\chi}^{i 2}\,  +\, v_\phi\, O_{\chi}^{i 1} \right) \label{eq:h2WW}
\end{eqnarray} 

\begin{eqnarray} 
\parbox{40mm}{ 
\begin{fmffile}{Diagrams/FeynDia59} 
\fmfframe(20,20)(20,20){ 
\begin{fmfgraph*}(75,75) 
\fmfleft{l1}
\fmfright{r1,r2}
\fmf{dashes}{l1,v1}
\fmf{wiggly}{r1,v1}
\fmf{wiggly}{r2,v1}
\fmflabel{$h_{{i}}$}{l1}
\fmflabel{$Z_{{\sigma}}$}{r1}
\fmflabel{$Z_{{\mu}}$}{r2}
\end{fmfgraph*}} 
\end{fmffile}}  
&=& iO_{h_i Z Z} g_{\sigma \mu}\nonumber \\
&=& \frac{i}{2}\, g_{\sigma \mu}\, v_\phi\, O_{\chi}^{i 1 *} \left(g_1 \sin\Theta_W\,   +\, g_2 \cos\Theta_W  \right)^{2} \label{eq:h2ZZ}
\end{eqnarray} 

\begin{eqnarray} 
\parbox{40mm}{ 
\begin{fmffile}{Diagrams/FeynDia10} 
\fmfframe(20,20)(20,20){ 
\begin{fmfgraph*}(75,75) 
\fmfleft{l1}
\fmfright{r1,r2}
\fmf{dashes}{l1,v1}
\fmf{dashes}{r1,v1}
\fmf{dashes}{r2,v1}
\fmflabel{$h_{{i}}$}{l1}
\fmflabel{$h_{{j}}$}{r1}
\fmflabel{$h_{{k}}$}{r2}
\end{fmfgraph*}} 
\end{fmffile}}  
 &=& i g_{h_i h_j h_k} \nonumber \\
 &=&\frac{i}{2} \left\{ O^{i 2}_{\chi} \left[-2 O^{{j 2}}_{\chi} \left(6 \lambda_{\phi\Omega} v_\Omega O^{{k 2}}_{\chi} + \lambda_{\phi\Omega} v_\phi O^{{k 1}}_{\chi} \right) \right. \right. \nonumber \\
 &+& \left. \left. O^{{j 1}}_{\chi} \left\{\left(-2 \lambda_{\phi\Omega} v_\Omega  + \sqrt{2} \mu_{\phi\Omega}  \right)O^{{k 1}}_{\chi}  -2 \lambda_{\phi\Omega} v_\phi O^{{k 2}}_{\chi} \right\}\right] \right. \nonumber \\ 
 &+& \left. O^{{i 1}}_{\chi} \left[O^{{j 2}}_{\chi} \left\{\left(-2 \lambda_{\phi\Omega} v_\Omega  + \sqrt{2} \mu_{\phi\Omega}  \right)O^{{k 1}}_{\chi}  -2 \lambda_{\phi\Omega} v_\phi O^{{k 2}}_{\chi} \right\} \right. \right. \nonumber \\ 
 &+& \left. \left. O^{{j 1}}_{\chi} \left\{\left(-2 \lambda_{\phi\Omega} v_\Omega  + \sqrt{2} \mu_{\phi\Omega} \right)O^{{k 2}}_{\chi}  -6 \lambda_\phi v_\phi O^{{k 1}}_{\chi} \right\}\right]\right\} \label{eq:h2h1h1}
\end{eqnarray} 

\begin{eqnarray}
\parbox{40mm}{ 
\begin{fmffile}{Diagrams/FeynDia111} 
\fmfframe(20,20)(20,20){ 
\begin{fmfgraph*}(75,75) 
\fmfleft{l1}
\fmfright{r1,r2}
\fmf{fermion}{v1,l1}
\fmf{fermion}{r1,v1}
\fmf{dashes}{r2,v1}
\fmflabel{$\bar{d}_{{i}}$}{l1}
\fmflabel{$d_{{j}}$}{r1}
\fmflabel{$h_{{k}}$}{r2}
\end{fmfgraph*}} 
\end{fmffile} }
 &=& i\left( A^{\overline{d}_i d_j h_k }_L\,P_L\,+\, A^{\overline{d}_i d_j h_k }_R\,P_R\right) \nonumber \\
 &=& -i \frac{1}{\sqrt{2}} O_{\chi}^{k 1}  U^{d *}_{(L)\,{j b}} U^{d *}_{(R)\,{i a}} Y_{(d)}^{{a b}}   P_L \nonumber \\ 
 &-&  i \frac{1}{\sqrt{2}} O_{\chi}^{k 1}  U_{(R)\,{j a}}^{d}   U_{L\,{i b}}^{d}  Y_{(d)}^{{a b}*} P_R
\end{eqnarray}

\begin{eqnarray} 
\parbox{40mm}{ 
\begin{fmffile}{Diagrams/FeynDia115} 
\fmfframe(20,20)(20,20){ 
\begin{fmfgraph*}(75,75) 
\fmfleft{l1}
\fmfright{r1,r2}
\fmf{fermion}{v1,l1}
\fmf{fermion}{r1,v1}
\fmf{dashes}{r2,v1}
\fmflabel{$\bar{u}_{{i}}$}{l1}
\fmflabel{$u_{{j}}$}{r1}
\fmflabel{$h_{{k}}$}{r2}
\end{fmfgraph*}} 
\end{fmffile}}
 &=& i\left( A^{\overline{u}_i u_j h_k }_L\,P_L\,+\, A^{\overline{u}_i u_j h_k }_R\,P_R\right) \nonumber \\
 &=& i \frac{1}{\sqrt{2}} O_{\chi}^{k 1}  U^{u *}_{(L)\,{j b}} U^{u *}_{(R)\,{i a}} Y_{(u)}^{{a b}}   P_L \nonumber \\ 
 &+&  i \frac{1}{\sqrt{2}} O_{\chi}^{k 1}  U_{(R)\,{j a}}^{u}  U_{(L)\,{i b}}^{u}        Y_{(u)}^{{a b}*} P_R
\end{eqnarray} 

\begin{eqnarray} 
\parbox{40mm}{ 
\begin{fmffile}{Diagrams/FeynDia114} 
\fmfframe(20,20)(20,20){ 
\begin{fmfgraph*}(75,75) 
\fmfleft{l1}
\fmfright{r1,r2}
\fmf{fermion}{v1,l1}
\fmf{fermion}{r1,v1}
\fmf{dashes}{r2,v1}
\fmflabel{$\bar{e}_{{i}}$}{l1}
\fmflabel{$e_{{j}}$}{r1}
\fmflabel{$h_{{k}}$}{r2}
\end{fmfgraph*}} 
\end{fmffile} }
&=& i\left( A^{\overline{e}_i e_j h_k }_L\,P_L\,+\, A^{\overline{e}_i e_j h_k }_R\,P_R\right) \nonumber \\
&=& -i \frac{1}{\sqrt{2}} O_{\chi}^{k 1} V^{e *}_{(L)\,{j b}} V^{e *}_{(R)\,{i a}} Y_{(e)}^{{a b}}   P_L \nonumber \\ 
&-&  i\frac{1}{\sqrt{2}}  O_{\chi}^{k 1} V_{(R)\,{j a}}^{e}   V_{(L)\,{i b}}^{e}   Y_{(e)}^{{a b}*} P_R
\end{eqnarray}  

\subsection{Diagrams Involving Charged Higgses}
\begin{eqnarray} 
\parbox{40mm}{ 
\begin{fmffile}{Diagrams/FeynDia55} 
\fmfframe(20,20)(20,20){ 
\begin{fmfgraph*}(75,75) 
\fmfleft{l1}
\fmfright{r1,r2} 
\fmf{dashes}{l1,v1}
\fmf{scalar}{v1,r1}
\fmf{boson}{v1,r2}
\fmflabel{$h_{{i}}$}{l1}
\fmflabel{$H^+_{{j}}$}{r1}
\fmflabel{$W^-_{{\mu}}$}{r2}
\end{fmfgraph*}} 
\end{fmffile}}
 &=& iO_{h_i H^+_{{j}} W^-}\Big(- p^{H^+_{{j}}}  + p^{h_{{i}}}\Big)_{\mu} \nonumber \\
 &=& \frac{i}{2} g_2 \Big(-2 O_{\chi}^{i 2} O_{c}^{j 2}  + O_{\chi}^{i 1} O_{c}^{j 1} \Big)\Big(- p^{H^+_{{j}}}  + p^{h_{{i}}}\Big)_{\mu} \label{eq:HpWh1}
\end{eqnarray} 

\begin{eqnarray} 
\parbox{40mm}{ 
\begin{fmffile}{Diagrams/FeynDia63} 
\fmfframe(20,20)(20,20){ 
\begin{fmfgraph*}(75,75) 
\fmfleft{l1}
\fmfright{r1,r2}
\fmf{scalar}{v1,l1}
\fmf{boson}{v1,r1}
\fmf{wiggly}{r2,v1}
\fmflabel{$H^+_{{i}}$}{l1}
\fmflabel{$W^-_{{\sigma}}$}{r1}
\fmflabel{$Z_{{\mu}}$}{r2}
\end{fmfgraph*}} 
\end{fmffile}}
 &=& iO_{H^+_{{j}} W^- Z}g_{\sigma \mu} \nonumber \\
 &=& -\frac{i}{2} g_{\sigma \mu} \left(2 g_2^2 v_\Omega O_{c}^{i 2} \cos\Theta_W   + g_1 g_2 v_\phi O_{c}^{i 1} \sin\Theta_W  \right) \label{eq:GmHpWZ}
\end{eqnarray} 

\begin{eqnarray} 
\parbox{40mm}{ 
\begin{fmffile}{Diagrams/FeynDia117} 
\fmfframe(20,20)(20,20){ 
\begin{fmfgraph*}(75,75) 
\fmfleft{l1}
\fmfright{r1,r2}
\fmf{fermion}{v1,l1}
\fmf{fermion}{r1,v1}
\fmf{scalar}{v1,r2}
\fmflabel{$\bar{u}_{{i}}$}{l1}
\fmflabel{$d_{{j}}$}{r1}
\fmflabel{$H^+_{{k}}$}{r2}
\end{fmfgraph*}} 
\end{fmffile}}
 &=& i\left( A^{\overline{u}_i d_j H_k^+ }_L\,P_L\,+\, A^{\overline{u}_i d_j H_k^+ }_R\,P_R\right) \nonumber \\
 &=& -i \left( U^{d,*}_{(L){j b}} U^{u,*}_{(R){i a}} Y_{u}^{a b} O^{{k 1}}_{\chi}P_L + U_{(R){j a}}^{d}  U_{(L){i b}}^{u}  Y_{d}^{a b *}  O^{{k 1}}_{\chi} P_R \right) 
\end{eqnarray} 

\begin{eqnarray} 
\parbox{40mm}{
\begin{fmffile}{Diagrams/FeynDia118} 
\fmfframe(20,20)(20,20){ 
\begin{fmfgraph*}(75,75) 
\fmfleft{l1}
\fmfright{r1,r2}
\fmf{plain}{l1,v1}
\fmf{fermion}{r1,v1}
\fmf{scalar}{v1,r2}
\fmflabel{$\nu_{{i}}$}{l1}
\fmflabel{$e_{{j}}$}{r1}
\fmflabel{$H^+_{{k}}$}{r2}
\end{fmfgraph*}} 
\end{fmffile}}
 &=& i\left( A^{\overline{\nu}_i e_j H_k^+ }_L\,P_L\,+\, A^{\overline{\nu}_i e_j H_k^+ }_R\,P_R\right) \nonumber \\
 &=& -i V_{(R){j a}}^{e}  V_{{i b}}^{\nu} Y_{e}^{a b*} O^{{k 1}}_{c} P_R
\end{eqnarray}

\newpage

\end{document}